\documentclass[amsmath,amssymb,prb,preprintnumbers,superscriptaddress,twocolumn]{revtex4}

\usepackage{graphicx}
\usepackage{dcolumn}    
\usepackage{bm}         
\usepackage{color}      
\usepackage{ulem}       

\newcommand{\lapprox}{{\scriptscriptstyle\stackrel{<}{\sim}}}

\begin{document}

\title{YBa$_2$Cu$_3$O$_7$/La$_{0.7}$Ca$_{0.3}$MnO$_3$ bilayers: Interface coupling and electric transport properties}

\author{R.~Werner}
\affiliation{%
  Physikalisches Institut -- Experimentalphysik II,
  Universit\"{a}t T\"{u}bingen,
  Auf der Morgenstelle 14,
  72076 T\"{u}bingen, Germany
}
\author{C.~Raisch}
\affiliation{%
  Physikalische Chemie,
  Universit\"{a}t T\"{u}bingen,
  Auf der Morgenstelle 18,
  72076 T\"{u}bingen, Germany
}
\author{A.~Ruosi}
\affiliation{%
  CNR-SPIN and Dept. of Physics, University of Naples Federico II,
  P.Tecchio 80,
  80125 Naples, Italy
}
\author{B.~A.~Davidson}
\affiliation{%
  CNR-IOM TASC National Laboratory,
  S.S. 14 Km 163.5 in AREA Science Park,
  34012 Basovizza, Trieste, Italy
}
\author{P.~Nagel}
\author{M.~Merz}
\author{S.~Schuppler}
\affiliation{%
  Forschungszentrum Karlsruhe,
  Institut f\"{u}r Festk\"{o}rperphysik,
  D-76021 Karlsruhe, Germany
}
\author{M.~Glaser}
\affiliation{%
  Physikalische Chemie,
  Universit\"{a}t T\"{u}bingen,
  Auf der Morgenstelle 18,
  72076 T\"{u}bingen, Germany
}
\author{J.~Fujii}
\affiliation{%
  CNR-IOM TASC National Laboratory,
  S.S. 14 Km 163.5 in AREA Science Park,
  34012 Basovizza, Trieste, Italy
}
\author{T.~Chass\'{e}}
\affiliation{%
  Physikalische Chemie,
  Universit\"{a}t T\"{u}bingen,
  Auf der Morgenstelle 18,
  72076 T\"{u}bingen, Germany
}
\author{R.~Kleiner}
\author{D.~Koelle}
\email{koelle@uni-tuebingen.de}
\affiliation{%
  Physikalisches Institut -- Experimentalphysik II,
  Universit\"{a}t T\"{u}bingen,
  Auf der Morgenstelle 14,
  72076 T\"{u}bingen, Germany
}
\date{\today}

\begin{abstract}
Heteroepitaxially grown bilayers of ferromagnetic La$_{0.7}$Ca$_{0.3}$MnO$_3$ (LCMO) on top of superconducting YBa$_2$Cu$_3$O$_7$ (YBCO) thin films were investigated by focusing on electric transport properties as well as on magnetism and orbital occupation at the interface.
Transport measurements on YBCO single layers and on YBCO/LCMO bilayers, with different YBCO thickness $d_Y$ and constant LCMO thickness $d_L=50$\,nm, show a significant reduction of the superconducting transition temperature $T_c$ only for $d_Y<10$\,nm,with only a slightly stronger $T_c$ suppression in the bilayers, as compared to the single layers.
X-ray magnetic circular dichroism (XMCD) measurements confirm recently published data of an induced magnetic moment on the interfacial Cu by the ferromagnetically ordered Mn ions, with antiparallel alignment between Cu and Mn moments.
However, we observe a significantely larger Cu moment than previously reported, indicating stronger coupling between Cu and Mn at the interface.
This in turn could result in an interface with lower transparency, and hence smaller spin diffusion length, that would explain our electric transport data, i.e.~smaller $T_c$ suppression.
Moreover, linear dichroism measurements did not show any evidence for orbital reconstruction at the interface, indicating that a large change in orbital occupancies through hybridization is not necessary to induce a measurable ferromagnetic moment on the Cu atoms.
\end{abstract}

\maketitle

\section{Introduction}
\label{sec:Introduction}

Singlet Superconductivity and ferromagnetism do not usually coexist in bulk compounds, as the exchange field in the ferromagnet favors an alignment of the conduction electron spins in the same direction, preventing the pairing effect in Cooper-pairs formed by electrons with anti-parallel spin.
However, the combination of superconducting (S) and ferromagnetic (F) materials in artificial thin layered systems, gives the unique opportunity to investigate the interplay between these two competing long-range order phenomena.
In such SF hybrid devices, superconducting correlations may be established in the ferromagnet due to the proximity effect, allowing superconductivity and ferromagnetism to coexist within a short distance from the interface of the order of the induced superconducting correlation length, $\xi_F$ \cite{Buzdin05, Bergeret04, Deutscher05}.
Simultaneously, the exchange field causes pair breaking in the superconductor, weakening or even suppressing the superconducting order parameter, and inducing a local magnetic moment in the superconductor at a distance from the SF interface set by the superconducting coherence length $\xi_s$.
Magnetic ordering is generally more robust than superconductivity (the exchange energy in ferromagnets is typically ~1 eV, while the Cooper pair formation energy is ~0.01 eV), and for materials having a strong exchange field, magnetism may be unperturbed by the proximity of a superconductor.

While SF hybrid structures based on metallic ferromagnets and conventional superconductors have been investigated in detail\cite{Buzdin05}, there are much less studies on SF systems involving high-transition temperature cuprate superconductors characterized by a very short coherence length and an anisotropic superconducting gap.
In this context, half metallic rare earth manganites like La$_{0.7}$M$_{0.3}$MnO$_3$ (M = Ca, Sr, Ba) are ideal ferromagnets, as they are nearly perfectly in-plane lattice matched with cuprates, which enables heteroepitaxial growth of cuprate/manganite SF bilayers and superlattices with well defined interfaces \cite{Varela03,Zhang09c}.
In particular, YBa$_2$Cu$_3$O$_7$/La$_{0.7}$Ca$_{0.3}$MnO$_3$ (YBCO/LCMO) superlattices have allowed the study of novel phenomena, such as a long range proximity effect\cite{Sefrioui03, Pena04}, spin polarized  quasiparticle injection into the S layer within a spin diffusion length $\xi_{\mathrm{FM}}$ \cite{Soltan04}, giant magnetoresistance\cite{Pena05} and a giant modulation of the F-layer magnetization induced by superconductivity\cite{Hoppler09}.

Recently, interfacial properties in YBCO/LCMO superlattices were investigated by X-ray magnetic circular dichroism (XMCD)\cite{Chakhalian06} and X-ray linear dichroism (XLD)\cite{Chakhalian07}.
These studies revealed an induced ferromagnetic moment on the interfacial Cu, oriented antiparallel to the adjacent Mn, whose temperature dependence follows that of the Mn moment.
The authors suggest that Cu and Mn are coupled across the interface by covalent chemical bonding that results in strong hybridization and large rearrangements of the orbital occupancies (orbital "reconstruction").
Within this context, the Mn-O-Cu superexchange interaction explains the induced magnetic moment in the cuprate and the presence of a non-superconducting YBCO layer at the interface.

Here, we present a detailed investigation of YBCO/LCMO bilayers, focusing on the dependence of transport properties on the YBCO layer thickness as well as on the interface coupling on an atomic length scale.
Transport measurements indicate high quality bilayers, showing a reduction of the superconducting transition temperature $T_c$ only below a YBCO thickness of $\approx 10$ unit cells.
Dichroism measurements using synchrotron radiation have been used to probe magnetic order and orbital occupations on both sides of the YBCO/LCMO interface by tuning the photon energy to Cu or Mn absorption resonances.
The XMCD measurements confirm the induction of a small net magnetic moment on Cu that vanishes near the Curie temperature of the LCMO.
Dichroism measurements with linearly polarized light show no evidence of any significant difference between the 3$d$ orbital occupations in the interfacial Cu as compared to the Cu in the bulk YBCO.
This implies that an induced magnetic moment on Cu through hybridization at the interface with Mn can result {\it without} any accompanying "orbital reconstruction".

\section{Experimental Details}
\label{sec:ExperimentalDetails}

Commercially available stoichiometric polycrystalline YBCO and LCMO targets were used for epitaxial growth of YBCO and LCMO thin films (with thickness $d_Y$ and $d_L$, respectively) and YBCO/LCMO bilayers by pulsed laser deposition on (001) SrTiO$_3$ (STO) substrates.
The targets were ablated by using a KrF ($\lambda$ = 248 nm) excimer laser at a repetition rate of $2\,$Hz\cite{Werner09}.
The substrate temperature $T_s$ during deposition was $750\,^{\circ}$C for all films for which data are presented below.
The oxygen pressure $p_{O_2}$ during thin film growth was 20\,Pa. After thin film deposition, the chamber was immediately vented with oxygen, and the samples were cooled down to $T_s=550\,^{\circ}$C in $p_{O_2}\approx1\,$mbar and annealed for $t=1\,$h to obtain fully oxidized films.
The cooling process was started right after the deposition to minimize interdiffusion at the bilayer interfaces.
For all bilayers shown here, YBCO was grown directly on STO and covered by LCMO.

In-situ high-pressure reflection high energy electron diffraction (RHEED) was used to monitor the growth mode and the exact number of deposited monolayers.
The surface morphology was checked by atomic force microscopy (AFM) in contact mode and the crystal structure was characterized by X-ray diffraction (XRD).
The thin film resistance $R$ was measured by a Van der Pauw method on unpatterned films in a temperature range of $T = 10-300$\,K in order to determine the superconducting transition temperature $T_c$ of the YBCO films or the metal-to-insulator transition temperature $T_{\mathrm{MI}}$ of the LCMO films.
Here, we define $T_{\mathrm{MI}}$ as the temperature for which $R(T)$ shows a maximum (which is typically within a few Kelvin of the ferromagnetic transition temperature $T_{\mathrm{Curie}}$ in LCMO .
A superconducting quantum interference device (SQUID) magnetometer was used to characterize the magnetic and superconducting properties of the samples by measuring magnetization $M(T)$ from $T=10-250\,$K in order to obtain $T_c$ of the YBCO films and $T_{\mathrm{Curie}}$ of the LCMO films.

In order to obtain site- and element-specific information regarding the local electronic structure (orbital occupation) and magnetic properties of YBCO/LCMO bilayers, we performed X-ray absorption spectroscopy (XAS), which is the absorption of an x-ray photon and the excitation of a core level electron into an unoccupied state through the electric dipole transition.
These experiments were performed at the high-energy branch of the advanced photoelectric effect (APE) beamline located at the ELETTRA storage ring in Trieste\cite{Panaccione09} and at the soft X-ray beamline WERA at the Angstr\"{o}mquelle Karlsruhe (ANKA).
All XAS data shown below (Secs.~\ref{sec:IF-spectroscopy}, \ref{sec:OrbitalOccupation}) were obtained from the same YBCO/LCMO bilayer and are representative of all samples measured.
This consists of a thin capping layer of 13 unit cell (uc) LCMO ($d_L\approx 5.2$\,nm), which was grown on top of a thicker layer of YBCO of about 18 uc ($d_Y = 20$\,nm).

The XAS data were recorded in surface sensitive total electron yield (TEY) mode and in bulk sensitive fluorescent yield (FY) mode.
In TEY mode, we probe predominantly interfacial Cu within the YBCO/LCMO bilayer.
Due to the small electron escape depth at the Mn or Cu L edge energies ($\approx 2$\,nm), most ($>90$\%) of the signal comes from within $6-8$\,nm of the surface (and is dominated by the $5$\,nm overlayer of LCMO).
Circularly polarized synchrotron radiation was used for measuring soft x-ray absorption spectra of the Cu and Mn L $_{\mathrm{2,3}}$ (2$p$ $\Rightarrow$ 3$d$ transition) absorption edges in TEY mode on the YBCO/LCMO bilayer.
At APE, a fixed photon-sample geometry was used (30$^{\circ}$ incident angle of light with respect to the film plane)  and polarization (circular, linear) was changed at the undulator located in the storage ring.
At WERA, the photon polarization was chosen by adjusting the exit slits after the bending magnet and was therefore fixed.
For the linear dichroism measurements, the electric field vector was aligned in the film plane or along the surface normal by changing the orientation of the sample normal.
The XAS spectra are normalized to equal step heights beyond ionization threshold.
XMCD spectra have been corrected for the incomplete photon polarization (90\,\%) and the 30$^{\circ}$ incident angle at APE.
Geometrical corrections for LD spectra taken at WERA have been made.
We define the XMCD signal (for a given energy) as the difference of the XAS signals (normalized to their average value) with incident light helicity oriented, respectively, parallel and antiparallel to the magnetization.
The XMCD measurements at APE were always measured in remanence, after applying around $\pm 60\,$mT in the film plane at 30$^{\circ}$ incidence angle with respect to the photon helicity.
We further recorded XLD data in both TEY and FY mode (in-plane or out-of-plane with respect to the sample surface).
The XLD signal is defined as out-of-plane minus in-plane XAS normalized to in-plane XAS intensity.

\section{Thin Film Characterization}
\label{sec:ThinFilmCharacterization}

\begin{figure}[b]
\centering
\includegraphics[width=0.93\columnwidth]{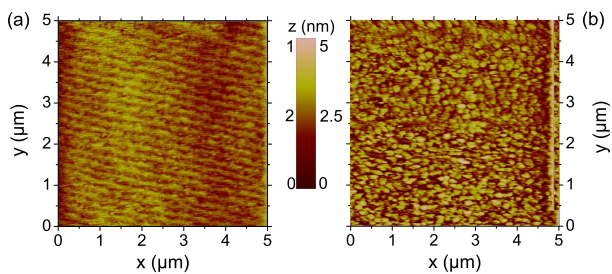}
\caption{(Color online)
AFM images of LCMO film surface ($d_L = 50$\,nm) with an rms roughness of $0.2$\,nm (a) and YBCO film surface ($d_Y = 50$\,nm) with an rms roughness of 0.8\,nm (b).
Numbers left and right from color bar refer to (a) and (b), respectively.}
\label{AFM}
\end{figure}

Figure \ref{AFM} shows typical AFM measurements for LCMO and YBCO films ($d_Y=d_L=50$\,nm).
In the case of LCMO [cf.~Fig.~\ref{AFM}(a)], the monolayer steps of the substrate are transmitted to the LCMO thin film.
The films are atomically flat with a root mean square (rms) roughness of $0.2$\,nm, determined over the scan area $5\times 5\,\mu{\rm m}^2$ shown in Fig.~\ref{AFM}.
The case of the YBCO [cf.~Fig.~\ref{AFM}(b)] is different, because YBCO grown on STO relaxes after a few monolayers, due to the larger lattice mismatch, and starts to grow in a 3D mode.
The rms roughness of the shown YBCO film is 0.8\,nm.
%

\begin{figure}[t]
\centering
\includegraphics[width=0.95\columnwidth]{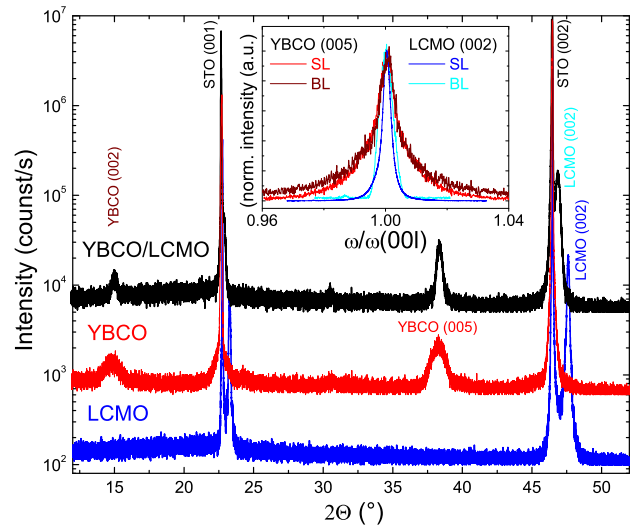}
\caption{(Color online)
XRD data for a single layer YBCO ($d_Y=10\,$nm) and LCMO ($d_L=50\,$nm) film and for a YBCO/LCMO bilayer ($d_Y=20\,$nm, $d_L=50\,$nm).
The main graph shows $\Theta-2\Theta$ scans (for YBCO and YBCO/LCMO shifted vertically for clarity).
The inset shows a comparison of rocking curves around the YBCO (005) and LCMO (002) peaks for the single layers (SL) and the bilayers (BL), with full width half maximum $\Delta\omega=0.05^\circ$ and $0.06^\circ$ for LCMO in the SL and BL and $0.10^\circ$ and $0.11^\circ$ for YBCO in the SL and BL, respectively.}
\label{XRD}
\end{figure}

Figure \ref{XRD} shows XRD data [$\Theta-2\Theta$ scans in the main graph and $\omega$ scans (rocking curves) in the inset] of a single layer YBCO ($d_Y=10\,$nm) and LCMO ($d_L=50\,$nm) film and a YBCO/LCMO (20\,nm/50\,nm) bilayer.
All samples are single phase and $c$-axis oriented.
Bulk YBCO has lattice constants $a=3.817\,${\AA},  $b=3.883\,${\AA} and $c=11.682\,${\AA}.
YBCO films grown on STO under optimized conditions relax their in-plane lattice constants within the first unit cells to the bulk values.
Bulk LCMO is orthorhombic, with pseudocubic lattice parameters $a=3.868\,${\AA}, $b=3.858\,${\AA} and $c=3.856\,${\AA}.
LCMO thin films grown on STO substrates or YBCO films grow fully strained for thicknesses up to $d_L=50\,$nm.
For LCMO films grown on STO this strain is tensile, so that the out-of-plane lattice constant of the LCMO is decreased; while LCMO grown on YBCO is under slight compressive strain.
Due to the different lattice mismatch between YBCO/LCMO and STO/LCMO, the LCMO (00$\ell$) peaks for the bilayer are shifted to smaller angles, as compared to the LCMO single layer, depending on the transmitted strain.
The low lattice mismatch between YBCO and LCMO (0.3\,\% in-plane), results in an excellent epitaxial growth of the bilayers.
This is confirmed by rocking curves around the (002) peak of LCMO and the (005) peak of YBCO, which yield almost the same values for single layer (SL) and bilayer (BL) films.
%

\section{Electric Transport Properties}
\label{sec:ElectricTransportProperties}

In this section we present and discuss results obtained on electric transport properties of YBCO/LCMO bilayers and compare those with the properties of single layer YBCO and LCMO films.

\subsection{YBCO and LCMO single layers}
\label{subsec:YBCO and LCMO single layers}

\begin{figure}[t]
\centering
\includegraphics[width=0.95\columnwidth]{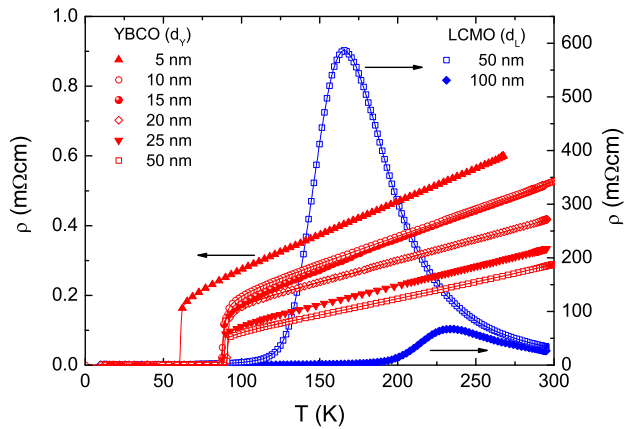}
\caption{(Color online)
Resistivity $\rho$ vs temperature $T$ of YBCO and LCMO single layer films with thicknesses $d_Y$ and $d_L$, respectively.}
\label{R(T)-SL}
\end{figure}

Figure \ref{R(T)-SL} shows resistivity $\rho$ vs temperature $T$ for six YBCO single layer films with $d_Y=5\ldots 50\,$nm and for two LCMO single layer films with $d_L=50$ and 100\,nm.
The metal-to-insulator transition temperature of the LCMO films is $T_{\mathrm{MI}} \approx 166$\,K (for $d_L=50\,$nm) and $\approx 235\,$K (for $d_L=100\,$nm).
The $T_\mathrm{MI}$ values depend on the oxygen content that influences the carrier density and on film strain that changes the strength of the double exchange interaction\cite{Salamon01}.
XRD reciprocal space-mapping (not shown here) shows that the 50\,nm LCMO film is coherently strained and therefore does not reach the bulk value of $T_{\mathrm{Curie}} = 260$\,K, while the 100\,nm LCMO film is relaxed and shows a much larger $T_\mathrm{MI}$ and a much smaller $\rho(T=T_\mathrm{MI})$.
The superconducting transition temperature $T_c$ of the YBCO thin films is $T_c \approx 85\ldots 90$\,K for all films, except for the thinnest one [for $T_c(d_Y)$ see Fig.~\ref{R(T)-Tc(d)}(b)], and the normal state resistivity $\rho$ increases with decreasing $d_Y$.
The room temperature resistivity is $\rho_L\approx 30$\,m$\Omega$cm and $\rho_Y\approx 0.3\ldots 0.6$\,m$\Omega$cm for the LCMO and YBCO films, respectively, i.~e.~$\rho$ differs by up to two orders of magnitude at room temperature.

\subsection{YBCO/LCMO bilayers}
\label{subsec:YBCO/LCMO bilayers}

\begin{figure}[t]
\centering
\includegraphics[width=0.95\columnwidth]{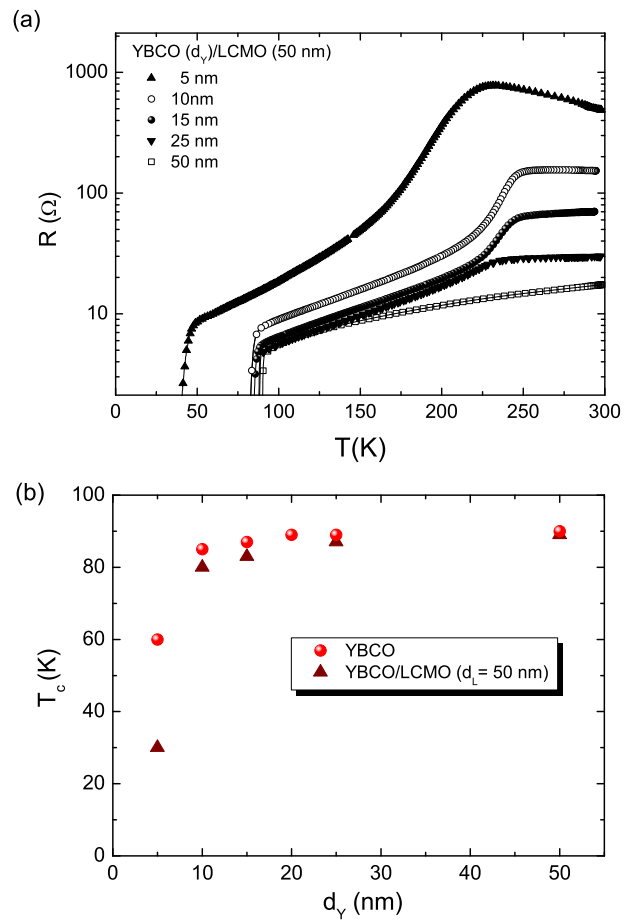}
\caption{(Color online)
(a) Resistance R vs temperature $T$ of YBCO/LCMO bilayers with $d_L=$50\,nm thick LCMO and different YBCO thickness $d_Y$.
(b) Superconducting transition temperature $T_c$ vs YBCO thickness $d_Y$ for YBCO single layers and for YBCO/LCMO bilayers obtained from the $R(T)$ data shown in (a) and in Fig.~\ref{R(T)-SL}.}
\label{R(T)-Tc(d)}
\end{figure}

We prepared bilayers with $d_L = 50$\,nm and $d_Y=5\ldots 50$\,nm; in all bilayers, YBCO was grown first and covered by LCMO.
The $R(T)$ dependence of five YBCO/LCMO bilayers with fixed $d_L=50\,$nm and variable $d_Y$ is shown in Fig.~\ref{R(T)-Tc(d)}(a).
The signature of $T_{\mathrm{MI}}$ is visible for all bilayers shown here, except for the one with the thickest YBCO layer.
While this is most clear for the sample with the thinnest YBCO layer, it gets less pronounced once $d_Y$ is increased, due to decreasing YBCO resistance.
Within the bilayers, $T_\mathrm{MI}$ of the LCMO film is significantly higher (around 230--240\,K) as compared to the single layer LCMO film with same thickness $d_L=50\,$nm
(cf.~Fig.~\ref{R(T)-SL}).
This can be attributed to the fact that LCMO grown on YBCO is much less strained as compared to the single layer LCMO film grown on STO \cite{Yang01}.

The suppression of $T_c$ with decreasing $d_Y$ is shown in Fig.~\ref{R(T)-Tc(d)}(b), both for YBCO single layers and YBCO/LCMO bilayers.
In both cases a significant suppression of $T_c$ is only observed for $d_Y=5\,$nm, and only for this smallest thickness we do observe a  clear difference in $T_c$ between single layer and bilayer samples.
We note that this observation is in contrast to Ref.~[\onlinecite{Soltan04}], where a drop in $T_c$$(d_Y)$ of YBCO/LCMO bilayers was found for $d_Y \lapprox$ 30\,nm.
The $T_c$ suppression observed for only very small $d_Y$ might indicate a smaller spin diffusion length of spin-polarized electrons into YBCO as compared to the one derived from $T_c(d_Y)$ data in Ref.~[\onlinecite{Soltan04}].
One possible explanation for this is a stronger interaction between the Cu and Mn moments at the interface for our samples.
The stronger hybridization could give rise to an electronically less-transparent interface (F/I/S, where I indicates insulating), that blocks injection of spin-polarized electrons.  To our knowledge, there has been no theoretical study of the electronic structure of the ferromagnetic YBCO layers for different strengths of hybridization.
In order to shed more light on this, we performed XAS measurements, which will be described below.
%

\section{Interface spectroscopy and proximity effect}
\label{sec:IF-spectroscopy}

\begin{figure}[t]
\centering
\includegraphics[width=0.95\columnwidth]{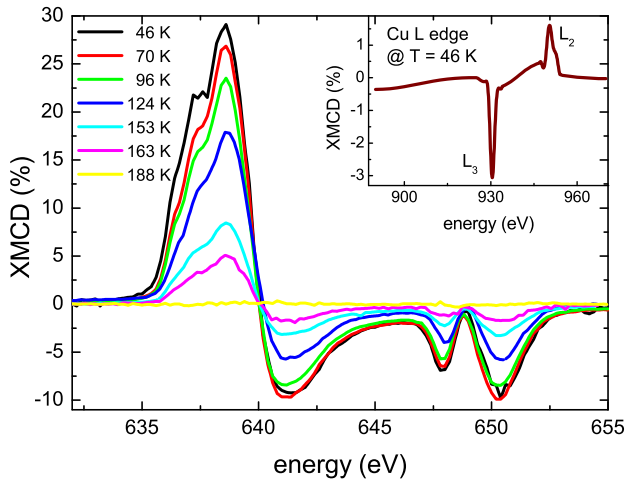}
\caption{(Color online)
XMCD spectra at the Mn L$_3$ edge for different $T$ from the YBCO/LCMO bilayer ($d_Y=20\,$nm; $d_L=5.2\,$nm).
The maximum dichroic signal is 29\,\% at low $T$ and decreases as $T$ approaches T$_{\mathrm{Curie}}$.
The inset shows the XMCD signal at the Cu L edge with a maximum of 3.0\,\% at $T=46$\,K, revealing antiferromagnetic coupling to the Mn magnetic moments.
As the Cu XAS signal is one order of magnitude smaller than for Mn, and as Cu dichroism is another order of magnitude smaller, we had to average over many scans and do a careful smoothing of the data.
Kinks in the signal arise from switching between different step sizes near the L$_3$ and L$_2$ edges.}
\label{XMCD}
\end{figure}

In the two following sections, we discuss XAS data obtained from the YBCO/LCMO bilayer with thicknesses $d_Y=20\,$nm and $d_L=5.2\,$nm.
Magnetization measurements $M(T)$
for this sample yielded $T_c\approx 80\,$K for the YBCO film and $T_{\mathrm{Curie}}\approx 200\,$K for the LCMO film within the bilayer.
Figure \ref{XMCD} shows XMCD spectra at the manganese L$_{2,3}$ edge ($\approx 640$\,eV) for different temperatures.
The shape of the spectra is typical for manganese atoms in a mixed Mn$^{3+}$/ Mn$^{4+}$ oxidation state \cite{Abbate92}.
The strong multiplet broadening of the Mn L$_3$ peak is a consequence of the partial occupation of the five Mn d orbitals.
The dichroism peak height at the L$_3$ edge (638 eV) at $T=46$\,K is 29\,\%.
The inset shows the XMCD signal at the Cu L edge with the L$_3$ and L$_2$ peaks at 931 eV and 951 eV, respectively, at the same temperature.
The non-zero XMCD signal indicates a ferromagnetic ordering of the Cu moments.
In addition, the opposite sign of Mn and Cu dichroism, reveals an antiparallel coupling between LCMO and YBCO across the heterostructure interface.

%

The maximum magnitude of the dichroism at the Cu L$_3$ edge at $T=46$\,K of about 3.0\,\% is higher than the value of 1.4\,\% reported in Ref.~[\onlinecite{Chakhalian06}] at $T= 30$\,K.
However, it is in good agreement with theory \cite{Yang09a}, predicting 2.4\,\% XMCD for YBCO/La$_{1-x}$Ca$_x$MnO$_3$ interfaces in which a single unit cell of YBCO is included in their model.
This demonstrates that the proximity to ferromagnetically ordered Mn spins induces spin canting in the Cu atoms of YBCO.
According to Ref.~[\onlinecite{Yang09a}], the presence of the ferromagnet leads to exchange splitting of the Cu d shell, resulting in spin-polarized states.
The hybridization at the interface of Cu $d_{3z^2}$ with spin-split Mn $d_{3z^2}$ states via O $p_z$ in the BaO layer (for an interface formed by adjacent layers of BaO and MnO$_2$ \cite{Varela03,Zhang09c})
then creates a slightly larger amount of holes in the majority than in the minority spin Cu $d_{3z^2}$ bands.
This produces a small net moment on the Cu sites.
Superexchange interactions determine the antiferromagnetic orientation of the Cu moment with respect to Mn.
In this model, the number of excess $d_{3z^2}$-derived majority states from the hybridization is small, sufficient for producing a measurable interfacial Cu XMCD signal, but not enough to modify the orbital occupancies that determine the experimentally measured linear dichroism.
This is in agreement with the LD measurements on our samples (as discussed in section \ref{sec:OrbitalOccupation}.)

\begin{figure}[b]
\centering
\includegraphics[width=0.95\columnwidth]{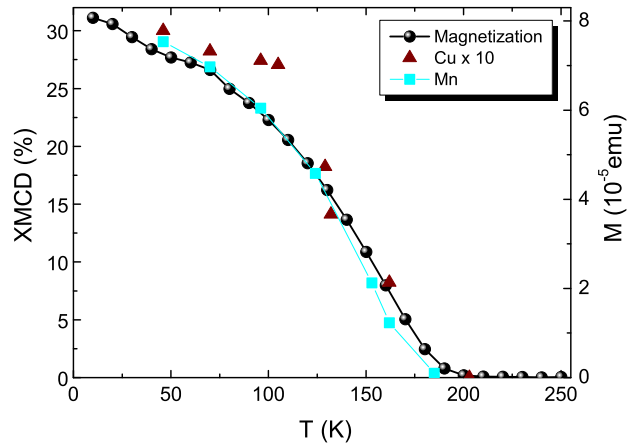}
\caption{(Color online)
Evolution of magnetic moments with temperature for YBCO/LCMO bilayer.
The comparison of XMCD signals (magnetic moments) at the Cu sites (triangles) and the Mn sites (squares) as well as bulk magnetization from SQUID measurements (black dots; field-cooled in 10\,mT) show the same behavior .
The Cu dichroism signal is scaled by a factor of 10.}
\label{XMCD(T)}
\end{figure}

Figure \ref{XMCD(T)} shows a comparison of the temperature dependence of XMCD signals and bulk magnetization $M$ obtained from SQUID measurements.
XMCD measured at the Mn L$_3$ edge is in good agreement with $M(T)$ from SQUID measurements, which yields $T_{\mathrm{Curie}} = 200$\,K.
The intensity of dichroism at the Cu L$_3$ edge (multiplied by a factor of 10) also decreases with increasing temperature until the signal becomes lower than a detectable threshold at $T=188$\,K.
The magnetic behavior of Cu closely follows the temperature dependence of the LCMO layer, persisting up to $T_{\mathrm{Curie}}$.
This supports the interpretation of induced ferromagnetism in YBCO across the YBCO/LCMO interface.
From sum rule calculations\cite{Chen95}, that relate the spin and orbital magnetic moments $m_S$ and m$_l$, respectively, to the areas of the L$_2$ and L$_3$ peaks, at low temperature we find $m_S=0.1\,\mu_B$/Cu to within an error of 20\,\%,
and $m_l < 0.03\,\mu_B$/Cu in remanence.
If the Cu moment is concentrated at the interface, the actual moment on the Cu atoms near the interface would be higher. Depending upon the assumed profile, this could imply an actual Cu moment higher by a factor of 2 or 3.
For a single hole 3d ground state (2p$^6$-3d$^9$) and a closed 3d shell final state (2p$^5$-3d$^{10}$) system these sum rule calculations are precise to within 5-10\,\% \cite{Piamonteze09}. Strong multiplet effects at the Mn L edge prevent a detailed analysis of manganese spin and orbital moments for this mixed Mn$^{3+}$/ Mn$^{4+}$ system.

\section{Orbital Occupation}
\label{sec:OrbitalOccupation}

\begin{figure}[b]
\centering
\includegraphics[width=0.95\columnwidth]{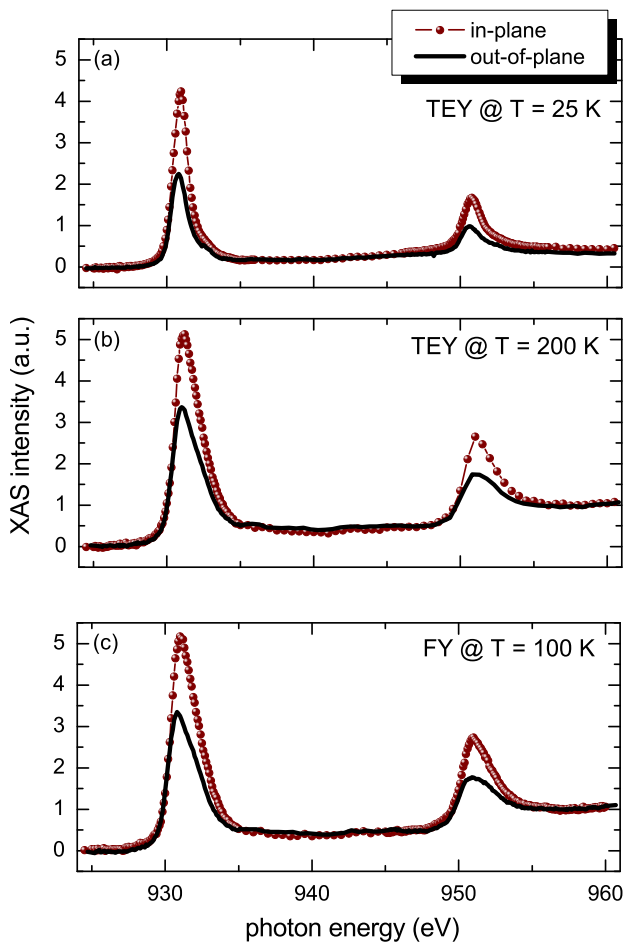}
\caption{(Color online)
Comparison of Cu L edge XA spectra taken with in-plane and out-of-plane polarization.
Shown are surface-sensitive TEY detection at $T=25\,$K (a) and  200\,K (b) and bulk-sensitive FY detection (c). Data in (a) was taken at ANKA; (b) and (c) at ELETTRA. }
\label{LD-Cu}
\end{figure}

In order to investigate the occupation of valence-electron orbitals on copper and manganese atoms we measured XLD in surface-sensitive TEY and bulk-sensitive FY detection mode on the YBCO/LCMO bilayer.
We were specifically interested in the orbital reconstruction proposed in Ref.~[\onlinecite{Chakhalian07}], where a large XLD signal in bulk YBCO but no XLD signal for YBCO in proximity to LCMO was reported.
The authors proposed orbital reconstruction to explain the almost identical occupation of in-plane and out-of-plane d-band states at the YBCO/LCMO interface.
However, in Ref.~[\onlinecite{Yang09a}] it is noted that for interfacial Cu the number of holes in the d$_{x^2-y^2}$ orbital is much bigger than the number of holes created in d$_{3z^2}$, which in turn would lead to negative linear dichroism at the interface.

\begin{figure}[b]
\centering
\includegraphics[width=0.95\columnwidth]{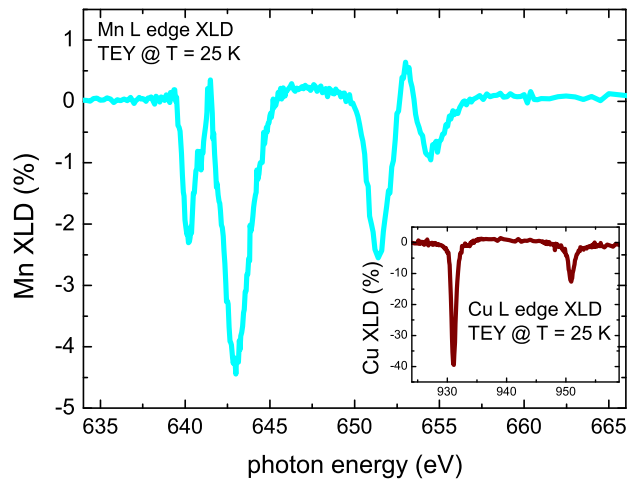}
\caption{(Color online)
XLD signal of the YBCO/LCMO bilayer at the Mn L edge in TEY mode at $T=25$\,K.
The inset shows the corresponding LD at the Cu L edge.}
\label{XLD}
\end{figure}

In fact we found a strong XLD signal at the Cu L edge [see Fig.~\ref{LD-Cu}(a) and inset in Fig.~\ref{XLD}] in TEY mode, 48\,\% at $T = 25$\,K at ANKA and 40\,\% at $T=46$\,K at ELETTRA (not shown here).
Moreover we do not observe any appreciable differences between bulk-sensitive FY detection [42\,\% XLD signal at $T=100$\,K, cf.~Fig.~\ref{LD-Cu}(c)] and interface-sensitive TEY detection [cf.~Fig.~\ref{LD-Cu}(b) and (c)], neither in shape nor in energy.
The shift of the XA edge energy of 0.4\,eV towards higher binding energy with increasing information depth, which was attributed to a charge transfer effect in Ref.~[\onlinecite{Chakhalian07}] could not be reproduced.
These results are evidence that it is possible to induce a magnetic moment on Cu across the interface without covalent bonding that drastically changes the orbital occupations.
The third piece of information obtained from linear dichroism is the temperature independence of orbital occupation.
Comparison between TEY data at $T= 25$\,K [Fig.~\ref{LD-Cu}(a)] and $T=200$\,K [Fig.~\ref{LD-Cu}(b)] show no difference in XLD signal.
This $T$ independence is robust, as expected for an energy level alignment by hybridization.
The orbital occupation being independent of temperature also emphasizes the absence of any signature on the Mn XMCD related to an induced superconductivity on interfacial LCMO.

Figure \ref{XLD} shows the XLD at the Mn L edge, which amounts to no more than 4.5\,\% (at the Mn L$_3$ edge).
The inset shows LD at the Cu L edge in interface sensitive TEY mode at $T = 25\,$K (see Fig.~\ref{LD-Cu} (a)).
In contrast to Chakhalian {\it et al.}\cite{Chakhalian07} who reported on the absence of XLD at the Mn L edge, stemming from an equal occupation of Mn d$_{x^2-y^2}$ and Mn d$_{3z^2}$ orbitals, we observe a negative XLD signal.
This is consistent with published reports of negative XLD of a few percent in very thin La$_{0.7}$Sr$_{0.3}$MnO$_3$ films grown on a number of different substrates \cite{Aruta06}.
Since the oxygen octahedra around Mn are only weakly distorted (lattice mismatch $< 0.3\,$\%), most of the linear dichroism should arise from interfacial Mn atoms.
Following the calculations of Ref.~[\onlinecite{Yang09a}], linear dichroism at the Cu and the Mn L edge should have the same sign at the YBCO/LCMO interface, because in both cases the density of unoccupied d$_{x^2-y^2}$ states is higher than for d$_{3z^2}$ states at the Fermi level.

\section{Conclusions}
\label{sec:Conclusions}

We examined bilayers of the high-temperature superconductor YBCO and the almost 100\,\% spin-polarized ferromagnet LCMO by means of electric transport measurements and {x-ray absorption spectroscopy.
The observation of a significant $T_c$ suppression only for very small YBCO thickness can be explained by the strong interaction between Mn and Cu moments at the interface, which we observed by XAS measurements.
Our XMCD data clearly confirm the} phenomenon of magnetic moments being induced on copper atoms at the LCMO/YBCO interface, with an even stronger interaction than found in the original report \cite{Chakhalian06}.
The effect is robust and closely follows the temperature dependence of magnetism in the manganite.
From the analysis of linear dichroism data, we conclude that covalent bonding and the resulting ''orbital reconstruction`` are not necessary for the spin canting of Cu moments in proximity to Mn spins.

\acknowledgments{R.~Werner gratefully acknowledges support by the Cusanuswerk, Bisch\"{o}fliche Studienf\"{o}rderung.
A.~R.~and B.~A.~D.~would like to acknowledge useful discussions with A.~Verna.
B.~A.~D.~acknowledges support by the FVG Regional project SPINOX funded by Legge Regionale 26/2005 and Decreto 2007/LAVFOR/1461.
ANKA Angstr\"{o}mquelle Karlsruhe and ELETTRA Synchrotron Trieste are acknowledged for the provision of beamtime.
This work was funded by the Deutsche Forschungsgemeinschaft (Project KL 930/11-2)}.


\end{document}